\documentclass[10pt,preprint2]{aastex}

\usepackage{natbib}
\usepackage{stfloats}

\shorttitle{Star Formation \& Extinction in the Disk of M82}
\shortauthors{Rodriguez-Merino et al.}

\begin{document}
\title{Spatially Resolved Star Formation History Along the Disk of\\ 
       M82 Using Multi-Band Photometric Data}

\author{L.\ H.\ Rodr\'{\i}guez-Merino, D. Rosa-Gonz\'alez \& Y. D. Mayya}

\affil{INAOE, Luis Enrique Erro 1, Tonantzintla, Puebla, Mexico, C.P. 72840.}

\email{lino@inaoep.mx}

\begin{abstract}
We present the results on the star formation history and extinction in the 
disk of M82 over spatial scales of 10$^{\prime\prime}$ ($\sim$180 pc). 
Multi-band photometric data covering from the far ultraviolet to the near 
infrared bands were fitted to a grid of synthetic spectral energy distributions. 
We obtained distribution functions of age and extinction for each of the 117 
apertures analyzed, taking into account observational errors through Monte-Carlo 
simulations. These distribution functions were fitted with gaussian functions 
to obtain the mean ages and extinctions along with errors on them. The analyzed 
zones include the high surface brightness complexes defined by \citet{OConnell78}.
We found that these complexes share the same star formation history and 
extinction as the field stellar populations in the disk. There is an 
indication that the stellar populations are marginally older at the outer disk 
(450~Myr at $\sim$3~kpc) as compared to the inner disk (100~Myr at 0.5~kpc). 
For the nuclear regions (radius less than 500~pc), we obtained an age of 
less than 10 Myr. The results obtained in this work are consistent with the idea 
that the 0.5-3~kpc part of the disk of M82 formed around 90\% of the 
stellar mass in a star-forming episode that started around 450~Myr ago lasting 
for about 350~Myr. We found that field stars are the major contributors to the 
flux over the spatial scales analyzed in this study, with stellar cluster 
contribution being 7\% in the nucleus and 0.7\% in the disk.

\end{abstract}


\section{Introduction}

The M82 galaxy (NGC 3034) is one of the most studied extra-galactic objects in 
the local universe. It is an edge-on spiral galaxy classified recently with a 
morphological type SBc \citep{Mayya05}. The small distance from the Milky Way 
to M82, 3.63 Mpc \citep[][]{Freedman94}, has allowed to obtain images with high 
spatial resolution, which were employed to study its strong star-forming activity 
\citep[][]{Solinger77,Telesco99,Forster01}, its compact star clusters (CSCs) 
distributed along the disk of the galaxy \citep[][]{OConnell78,OConnell95,DeGrijs01,
Melo05,Smith06,Smith07,Mayya08,Konstan09}, its biconical outflow which produces a 
filamentary structure of several kiloparsec-long \citep[][]{Lynds63,Cappi99,Hoopes05}, 
and the properties of the neutral and molecular gas \citep[][]{Shen95,Neininger98,
Matsushita05}. Particularly, the observed bridge of neutral hydrogen, which is 
connecting M82 with M81, has been used as a test of the interaction between these 
objects ~\citep[][]{Cottrell77,Yun93,Sofue98,Chynoweth08}. There were attempts 
to determine the exact epoch of interaction, with the estimated epoch depending 
highly on the assumptions in the simulations. \citet{Yun93} reproduced the observed 
HI plumes around M82 by a model where the last passage of the companion occurred 
around 300~Myr ago. \citet{Sofue98} estimated the epoch of interaction as around 
1~Gyr ago. This was obtained assuming that the fundamental stellar disk of M82 was 
blown away during the interaction, and that is the reason for the observed nearly 
Keplerian rotation curve. Given that the disk is rich of a population of AGB stars 
\citep{Davidge08} --- the products of post-interaction star-formation episode --- 
the timescale seems to be longer than the 300~Myr estimated by \citet{Yun93}. 
We hence use an intermediate age of 500~Myr as the epoch of interaction, for 
the sake of discussion in this work.

\citet{OConnell78} using the best available photographic images of M82 at that time 
analyzed the brightest areas of M82. They identified 8 complexes, which they 
denoted by the letters A to H, and pointed out that most of these complexes 
contain knotty structures. They found that along the disk (r $\ge$1 kpc) there is 
no trace of complexes with on-going star formation. They also noted that the nuclear 
part of the galaxy (radius less than 500 pc) presents a strong, recent star formation 
activity, with the complex A being the most intense. \citet{Rieke80,Rieke93} 
analyzed the multi-band properties of this zone, and determined an age of less than 
30~Myr for the nuclear region. \citet{Forster01} carried out a spatially resolved 
study of stellar ages in this region, concluding that the nuclear zone had more than 
one burst of star formation in the last 10~Myr. These authors have noted that the 
patchy appearance of the nuclear zone is due to dust obscuration with more than 4 mag 
of visual extinction if the majority of the obscuring dust resides in a foreground 
screen and 43--52 mag if dust is mixed with the stars. Sub-arcsec spatial resolution 
images offered by the {\it Hubble Space Telescope} (HST) allowed \citet{OConnell95} to 
discover more than a hundred compact stellar clusters in the nuclear zone. \citet{Smith06}, 
using HST/STIS spectroscopy, found that these clusters are young and are formed as 
part of the nuclear starburst that have been taking place over the last 10~Myr.

The complexes outside the nuclear zone are not as extensively studied as those in 
the starburst zone. \citet{DeGrijs01}, using the HST/WFPC2 images, found plenty of 
compact stellar clusters in the complex B, all older than those in the nuclear region, 
but younger than around 500~Myr. They found that the star formation rate in complex 
B was as high as the nuclear complex at the present day, thus giving it the name of 
fossil starburst. More recently, \citet{Smith07} has derived a mean age of around 
150~Myr for 35 clusters in the complex B. Complex F is the only other zone where 
studies in some detail have been carried out \citep{Smith01,Bastian07}. These studies 
gave an age of less than 100~Myr for this complex, with no evidence for recent star 
formation. More recently, \citet{Konstan09} have derived spectroscopic ages of around 
50 compact star clusters distributed over the disk, with the median age of the clusters 
being 150~Myr.

In order to explain the presence of patchy bright complexes in the nucleus and the 
disk, \citet{OConnell78} invoked a model where the star formation was restricted to 
only the bright complexes. They suggested that after the tidal encounter of M82 with 
the galaxy M81, some interstellar material was stripped from M82. When this material 
fell back into the galaxy, it produced the classical bright complexes with enhanced 
episodes of star formation. If this model is correct then the main population of stars 
in the disk surrounding the bright zones should be older than the bright zones 
themselves.

On the other hand, \citet{Mayya06} using photometric, dynamical and chemical 
properties of the disk between 1 and 2.7~kpc radius proposed a violent star-formation 
episode throughout the disk of M82 that started around 0.8~Gyr ago,and lasted for 
a few hundred million years. This model of star formation reproduces well the 
relative strengths of age-sensitive spectroscopic features in the blue part of 
long-slit spectra along the major axis, as well as the observed value of M/$L_k$, 
metallicity and enrichment of $\alpha$-elements. They also noted that 
the galaxy does not possess a massive old underlying stellar disk, characteristic 
of disk galaxies. More than 90\% of the stellar mass in the disk was formed in this 
episode, implying that the star-formation episode is not just restricted to a few 
patches as was suggested by \citet{OConnell78}, instead it was disk-wide. The basic 
tenets of this scenario were recently tested by \citet{Davidge08} using near infrared 
color-magnitude diagrams (CMDs) of the resolved populations in the disk. He detected 
asymptotic giant branch stars (AGBs) of age of a few hundred million years which 
are distributed uniformly in the disk, up to a radial distance as large as 12~kpc, 
suggesting that the star formation episode pervaded the entire disk. More importantly, 
red super giants (RSGs) were absent in the disk outside a radius of 0.5~kpc, confirming 
the cessation of star-formation episode $\sim100$~Myr ago.

While the bursty nature of star-formation history in the disk of M82 is by now 
well established, there is still room for improvement in the timing of the commencement 
of the disk-wide star-formation. In the study of \citet{Mayya06}, the uncertainty 
in age estimation comes from the errors involved in determining the stellar mass of 
the disk (to calculate the quantity M/$L_k$), which was obtained using a dynamical 
model to fit the observed rotation curve of the disk. Given that M82 has suffered a 
major change dynamically as a result of its interaction with the members of M81 group, 
and the evidence for the existence of gas above the plane, some of the intrinsic 
assumptions made to determine the mass, such as planar circular motions, may not be 
completely valid. An underestimation of the mass would push the commencement age to 
be as old as 1~Gyr, whereas if the dynamical masses were over-estimated, the mean 
ages could be younger than 0.5~Gyr. On the one hand, the analysis of the AGB stars 
carried out by \citet{Davidge08} promises to provide an accurate clock for age 
dating. However at present its utility is limited because of the poor knowledge of 
this evolutionary phase. Most recent models of \citet[][]{Marigo08} indicate age 
$>$200-300~Myr, with peak of the AGB population occurring at age $>$500~Myr. Hence 
alternative techniques are necessary to determine the age of the stellar disk in M82.

In recent years, the SEDs formed from photometric data covering a wide wavelength 
interval are increasingly used in determining the age, extinction and metallicity of 
stellar systems. \citet[][]{Bianchi05} using ultraviolet (UV) images from {\it Galaxy 
Evolution Explorer} (GALEX) and optical images from the {\it Sloan Digital Sky Survey} 
(SDSS), derived ages and reddening (among other parameters) of 160~pc scale stellar 
complexes in M101 and M51. \citet[][]{Kaviraj07} estimated age and metallicity of 
42 low-reddening globular clusters in M31 using the photometry in the FUV, NUV, U, B, V, 
R, I bands. \citet[][]{Bridzius08} explored the capability of the UBVRIJHK photometric 
system to estimate the age, metallicity and color excess of star clusters. These 
studies have illustrated that the long base line in wavelength helps to disentangle the 
effects of reddening from the evolution. SEDs can also be used to get a measure of the 
metallicities in old, low-reddening systems such as globular clusters. However, the 
photometrically-derived metallicities do not reach the accuracies that are achievable 
using spectroscopic data.

In the present study, we exploit the potential of present-day SED-fitting techniques 
to determine the ages of the dominant stellar populations in the disk of M82, and in 
the process spatially map the star formation history. The available photometric data 
allow us to analyze the star formation history over spatial scales of $\sim180$~pc, 
the size limitation basically coming from the 5$^{\prime\prime}$ beam of the GALEX 
images and the requirement that measured fluxes in each aperture have the enough signal 
to noise (S/N) ratio. In earlier studies \citep{Mayya06,Davidge08}, star formation 
history was inferred by analyzing data over spatial scales of 1--2~kpc, so the present 
study represents an improvement by a factor of around 10 in spatial resolution. 
Typically, the complexes defined by \citet{OConnell78} have spatial extensions comparable 
to our resolution element, with the exception of complex B, which is several times 
larger than our aperture. The use of imaging data to determine ages also allows a comparison 
of star formation history of the bright complexes with those of the surrounding disk.

In~\S 2, we describe the observational data used in this work. In~\S 3, we describe 
the method used to infer the age of the stellar population and extinction. In~\S 4, 
we describe the results obtained for the named complexes. In~\S 5, we analyze 
the spatial distribution of the stellar population age to infer the star formation 
history in the disk and nucleus of M82. In~\S 6, we determine the cluster formation 
efficiency and in~\S 7, we discuss our results. Concluding remarks are given in~\S 8.

\begin{figure*}[t]
\begin{center}
\includegraphics[scale=0.88]{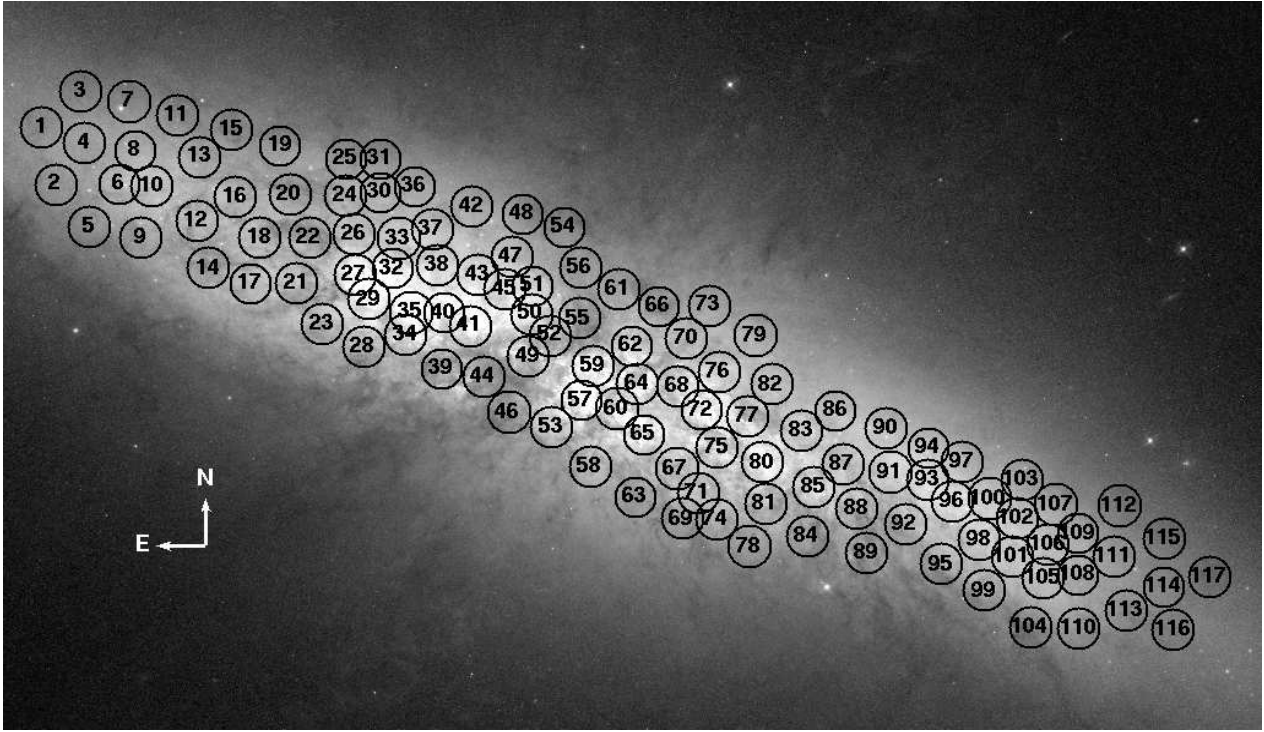}
\caption{F555W image of M82 from {\it HST}. The positions of the 117 apertures 
         (5$^{\prime\prime}$ radius) defined in this work are marked with circles. 
         The classical M82 complexes A, C, D, E, F, G, and H correspond to 
         the apertures 57, 65, 59, 64, 72, 80 and 52 respectively. The complex B 
         contains the apertures 27, 29, 32, 35, 40, 41, 45 and 50.} 
\label{fig:apertures}
\end{center}
\end{figure*}

\section{Observed Spectral Energy Distributions}

The most important part for deriving the spatially resolved star formation
history from photometric data is the availability of high S/N ratio images 
covering the wavelength interval from the far ultraviolet (FUV) to the 
near infrared (NIR). The S/N ratio in the analyzed zone should be better 
than 20 in order to obtain reliable ages.

The UV part of the SEDs is based on the GALEX far-UV (FUV) and near-UV (NUV) 
images \citep{Morrissey07}. The optical images in the u, g, r, i and z bands 
were retrieved from the SDSS database \citep{Abazajian03}. For the near infrared 
data, we used the J, H and K images obtained with the Cananea Near-Infrared 
Camera (CANICA) attached to the 2.1 m telescope at Observatorio {\it Astron\'omico 
Guillermo Haro} \citep[OAGH, ][]{Mayya05}. 

Among all these dataset, the {\it GALEX} images have the poorest spatial resolution 
with a point spread function of full width half maximum of 5$^{\prime\prime}$. 
These images set the minimum spatial scale of our analysis, which corresponds to 
$\sim$100 pc in M82. We carried out photometry in fixed apertures of 5$^{\prime\prime}$ 
radius. The apertures were placed so as to sample the bright complexes defined 
by \citet{OConnell78}, as well as the diffuse part of the disk. Figure 
\ref{fig:apertures} displays the F555W image of M82 from HST, showing the positions 
of the 117 selected apertures. In Table \ref{tab:regs}, we show the association of 
our aperture numbers and \citet{OConnell78} complexes. Aperture numbers increase 
from east to west along the major axis of the galaxy. Apertures  50 to 80 (see 
Figure~\ref{fig:apertures}) lie within the 500 pc of the center of the M82 and belong 
to the nuclear starburst.

\begin{figure}[t]
\epsscale{1.0}
\plotone{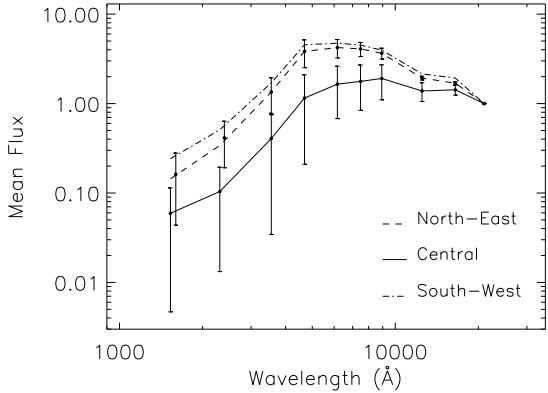}
\caption{Mean SEDs across the M82 disk. Solid line represents the mean of SEDs for selected 
         nuclear apertures. Dashed and dot--dashed lines are the mean SEDs for apertures 
         located in the north-east and south-west of M82 respectively. The mean SEDs were 
         normalized in the K-band flux and the vertical bars represent the flux dispersion 
         (1$\sigma$).}
\label{fig:seds}
\end{figure}

We aimed that the ages inferred from the SED-fitting technique are good to 15\%. 
We found that the photometric accuracy of each band used for the construction of 
the SED should be better than 5\% in order to achieve this. This corresponds to 
limiting surface brightnesses (for a S/N ratio of 20), averaged over an aperture 
of 5$^{\prime\prime}$ radius, of: FUV=28.01, NUV=26.85, u=19.34, g=19.97, r=18.66, 
i= 18.23, z= 16.69 (all in units of AB mag\,arcsec$^{-2}$), J=23.16 (Vega), H=19.89 
(Vega) and K=19.23 (Vega). M82 is brighter than these limits up to a galactocentric 
radius of 3~kpc in all bands, except for 8 apertures located in dusty features, 
whose $FUV$ and $u$ magnitudes were fainter than the limiting magnitudes. We 
eliminated these apertures from rest of the analysis. Once the apertures were defined, 
photometry is carried out by summing all the sky-subtracted flux inside the apertures.
The sky value for an image is obtained as the mean value of several selected areas 
outside the main disk of M82. Obvious sources, such as stars and background objects 
were avoided in the definition of sky zones. Note that we subtract sky, not the 
background, thus ensuring that our photometry measures the flux emitted by all the 
stars in the disk.

The error ($\Delta I$) in the flux ($I$) within the aperture was calculated 
by using,

\begin{equation}
\frac {\Delta I}{I} = \frac {\sqrt{\frac{counts+sky}{gain} + Npix*\sigma_{BG}^2}}{counts}\;,
\end{equation}
where, {\it Npix} is the number of pixels inside the aperture, {\it counts} and {\it sky} 
are the total number of counts in selected areas in the disk, and in areas free of sources, 
respectively. The $\sigma_{BG}$ in the equation is the RMS error per pixel, which was 
estimated using the sky regions in each image, and {\it gain} is  the number of electrons 
per data counts which is provided by the {\it GALEX} and {\it CANICA} web pages, and in 
the case of {\it SDSS}, it is found in the related {\it fpAtlas} file. The instrumental 
errors in the estimated aperture fluxes calculated using the above formula is less than 
1\% in all bands except in the bluest bands ($FUV$, $NUV$ and $u$). In the $FUV$-band the 
errors for majority of the apertures lie between 5--10\%,while errors lie between 1--5\% 
for majority of the apertures in the $NUV$ and $u$ bands. On the other hand, external errors, 
which are mainly determined by the errors in flux calibration between the three datasets, 
could be as large as 5\%. Both errors were added in quadrature to estimate the total error 
on each datapoint. 

In each image, we obtained the sky-subtracted fluxes in the selected apertures to 
construct 117 spectral energy distributions (SEDs). Figure \ref{fig:seds} shows the mean 
of nuclear SEDs (solid line), north-east SEDs (dashed line), and south-west SEDs (dot-dashed 
line). The SEDs were normalized to the K-band flux, to facilitate the comparison. The 
dispersion of individual SEDs from the mean is shown by the vertical bars at each filter 
band. It can be seen that the SEDs of north-east and south-west follow a similar trend 
with the latter being marginally bluer. The behavior of the mean SED of the central apertures 
is different. It is redder and peaks at longer wavelengths as compared to the apertures in 
the disk. Understanding these differences in terms of different star forming history and 
extinctions, is one of the topics of discussion in the current paper.

\begin{figure*}[t]
\epsscale{1.5}
\plotone{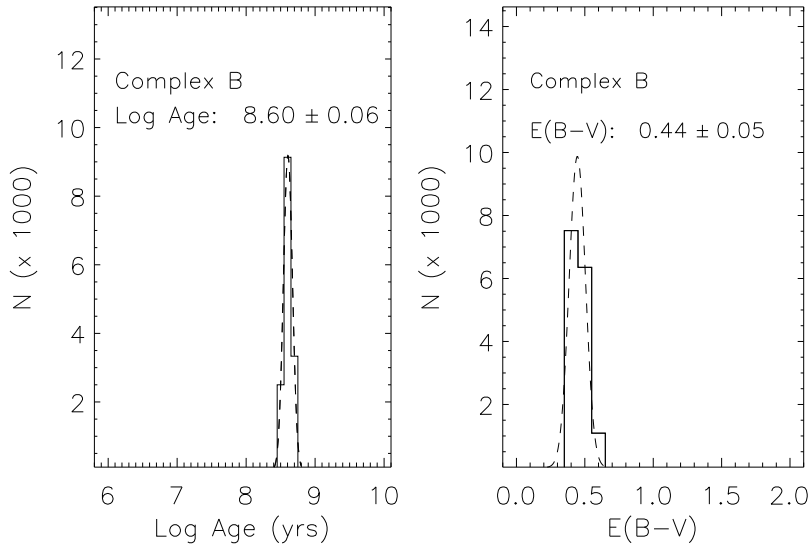}
\caption{Age and color excess obtained for complex B (aperture 27). The Monte-Carlo 
         simulation provides several best-fit models. The fitted gaussian (dashed 
         line) gives us the value of the age (left) and extinction (right) and their 
         corresponding errors.} 
\label{fig:bestfit}
\end{figure*}

\section{The SED Fitting Method}

The method we followed to determine the age and extinction of the stellar population 
has six steps:

\begin{enumerate}
\item Construction of synthetic SEDs:\\
The first step in obtaining the age and extinction of the stellar population
is the construction of synthetic SEDs, suitable for comparison with the
observed SEDs. 
      We used  a set of 99 simple stellar population (SSP) models computed
      by the Padova Group \citep{Bressan93} \footnote{most recently 
      updated models kindly provided by A. Bressan.}.  
      The set of models cover evenly -- in logarithmic scale -- an age range  
      from 1 Myr to 10 Gyr. These models use Salpeter's IMF and have solar metallicity.
      \citet{McLeod93} using nebular lines found that the metallicities of 
      the nuclear regions of M82 are close to solar. There are no metallicity measurements
      in the disk regions due to the lack of emission lines there. Following the
      observed trend of metallicities staying constant or decreasing with galactocentric 
      distance in late-type galaxies, we expect solar or sub-solar metallicities 
      in the disk. However, we used solar metallicity models for all the regions.

      In order to compare 
      synthetic SSP spectra with the observed SEDs, we construct a grid of synthetic 
      colors by integrating each synthetic spectrum over the response curve of 
      the filters used in the observations described in the previous section.

\item Reddening the synthetic SEDs:\\
      The synthetic SEDs were reddened using the \citet{Calzetti94} extinction 
      law. We also explored the results obtained if we make use of the Cardelli 
      law \citep{Cardelli89}. 
      Once the relative reddening is fixed by assuming an extinction curve, the 
      total extinction is computed by varying the color excess, E(B--V), from 0.0 
      to 2.0~mag, in steps of 0.1~mag. The number of synthetic reddened SEDs 
      is around 2000.

\item Definition of the merit functions:\\
      The best fits between observed and reddened synthetic SEDs in each aperture 
      were found by using two merit functions. The first merit function compares 
      the observed colors with the reddened SSP colors, and was defined as,

      \begin{equation}
      \chi^2 = \frac{\sum_{i} \frac{w_{i}\;(color_{i}^{\rm obs}-color_{i}^{\rm mdl})^2}
      {(\sigma_{i}^{\rm obs})^2}}{\sum_{i} w_{i}} \;,
      \end{equation}

      where the index $i$ runs for 9 different colors, using as a reference the $r$ 
      band  (FUV-{\it r}, NUV-{\it r}, {\it u-r}, {\it g-r}, {\it r-i}, {\it r-z}, 
      {\it r}-J, {\it r}-H and {\it r}-K). The variables $color_{i}^{\rm obs}$ and 
      $color_{i}^{\rm mdl}$ are the observed and synthetic colors respectively, and 
      $\sigma_{i}^{\rm obs}$ are the observational errors (instrumental and external 
      errors), the  weight of each individual color is given by $w_i$. 
      Some of the color indices are more sensitive to age and reddening than 
      others, which could, in principle, be reflected by the corresponding weights.
      However, after several trials, we found that the best way to get a good 
      fit (low-$\chi^2$) is to give the same weight for all the colors and introduce a 
      separate merit function to break the age-reddening degeneracy.

      The {\it u-g} color is chosen as a second merit function,
      given that the Balmer jump is the best discriminator 
      between populations of different ages, for ages less than around 500~Myr, 
      and that it is least sensitive to extinction. It is defined as: 

      \begin{equation}
         \beta^2= \chi^2 \; [(u-g)^{obs}-(u-g)^{mdl}]^2\;.
      \end{equation}

\item Selection of best-fit models:\\
      Due to the age-extinction degeneracy, different combinations of age and color 
      excess give us similar SEDs and therefore similar values of $\chi^2$ and $\beta^2$. 
      Instead of the common practice of retaining only the minimum $\chi^2$ model, 
      we retained 50 SSPs (out of the 2000 reddened SSPs) with the lowest $\chi^2$.
      Of these 50 SSPs, we then retained only those fits that
      are within 10\% of the minimum value of $\beta^2$. Thus, for every aperture, we
      have a minimum of 1 and a maximum of 50 best-fit solutions.

\item Monte-Carlo simulation of the observational errors:\\ 
      The error in each photometric band of an observed SED allows a set of 
      synthetic SEDs to reproduce the observed SED with similar value of $\chi^2$. 
      In order to select all the 
      synthetic SEDs that can fit the observed SED of an aperture, we generated
      5000 new SEDs, where the flux in each photometric band was re-calculated 
      by adding an error to the observed flux, where the error is randomly
      chosen such that their distribution is gaussian with a sigma equal to 
      $\sigma_{i}^{\rm obs}$ for each band. We refer this set of new SEDs as 
      Monte-Carlo SEDs or for brevity {\it MC-SEDs}. We repeat steps 3 and 4 
      with each of the 5000 {\it MC-SEDs}, and obtain the set of best-fit models 
      (maximum of 50 models) for each one, resulting in a large set (several 
      thousands) of best-fit combinations of ages and extinctions for each aperture.

\item Determination of age and extinction:\\
      The age and extinction associated with each of the thousands of best-fit 
      SSPs were analyzed statistically using a gaussian distribution function, 
      to determine the most likely age and extinction of a region. Gaussian functions 
      are fitted to the distributions of $\log({\rm age})$ and $E(B-V)$, with the 
      center of the gaussians giving the most likely value, and sigma, an estimation 
      of the error on the determined value. In a few cases where the distribution 
      has more than one peak, the gaussian function is fitted to the group having 
      the maximum number of best-fits. The method is illustrated in Figure 
      \ref{fig:bestfit} for the brightest aperture in complex B.

\end{enumerate}

\begin{figure}[t]
\begin{center}
\epsscale{1.05}
\plotone{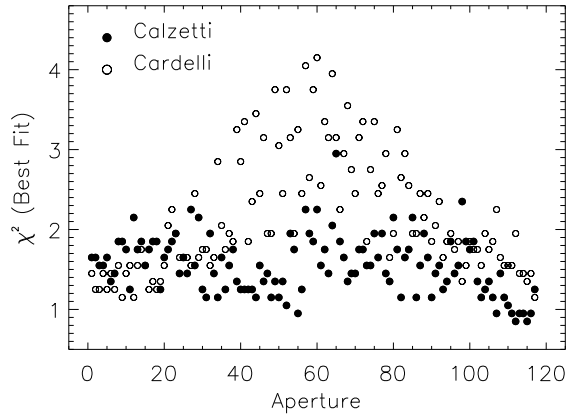}
\caption{The minimum $\chi^2$ values obtained using Calzetti (solid circles) and Cardelli 
         (open circles) extinction curves for all the selected apertures. Calzetti attenuation 
         curve provides lower $\chi^2$ values, especially in the dusty central zone (apertures 
         30-90).}
\label{fig:chi2_min}
\end{center}
\end{figure}

We found that the width of the distribution of ages and extinctions for a given 
region (defined by the $\sigma$ of the best-fit gaussian) is within 15\% of the derived 
value.
The uncertainties were $>$30\% if the observational errors in any of the bands is 
greater than 10\%. As mentioned before, we discarded eight apertures based on that 
restriction. All these discarded apertures have FUV magnitudes fainter than 
the $20\sigma$ limiting magnitude.

\begin{figure}[t]
\begin{center}
\epsscale{1.05}
\plotone{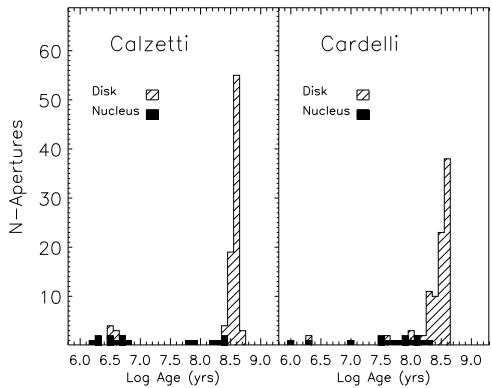}
\caption{Ages of the apertures obtained using Calzetti (left) and Cardelli (right) attenuation 
         curves. Both the curves give similar ages for disk, where as Cardelli curve gives 
         systematically higher ages ($\sim10^8$ yr instead of $\sim10^6$ yr ) for nuclear apertures.}
\label{fig:ages_calzetti_cardelli}
\end{center}
\end{figure}

We  adopted the Calzetti extinction curve in our analysis. However, we have investigated 
the effect of using the \citet{Cardelli89} extinction curve instead. The $\chi^2$ values 
for the best-fit model for each aperture using the two different extinction curves are 
compared in Figure~\ref{fig:chi2_min}. Both extinction curves give similar values of 
$\chi^2$ for the disk apertures where the visual extinction is less than 2 mag. 
For the nuclear apertures Calzetti law gives systematically better fits. The ages 
derived for the disk apertures using the two curves do not differ much as can be seen 
in Figure~\ref{fig:ages_calzetti_cardelli}. However we found that the ages derived 
using the Cardelli curve for majority of the nuclear regions are around $10^8$ years, 
which is clearly inconsistent with the estimation of ages using other methods 
\citep[e.g.][]{Forster03}. Hence, we use the results obtained with the Calzetti curve 
throughout this paper.

\section{Analysis of Derived Ages and Extinctions in Named Complexes}

\begin{figure*}[t]
\begin{center}
\includegraphics[width=0.6\textwidth]{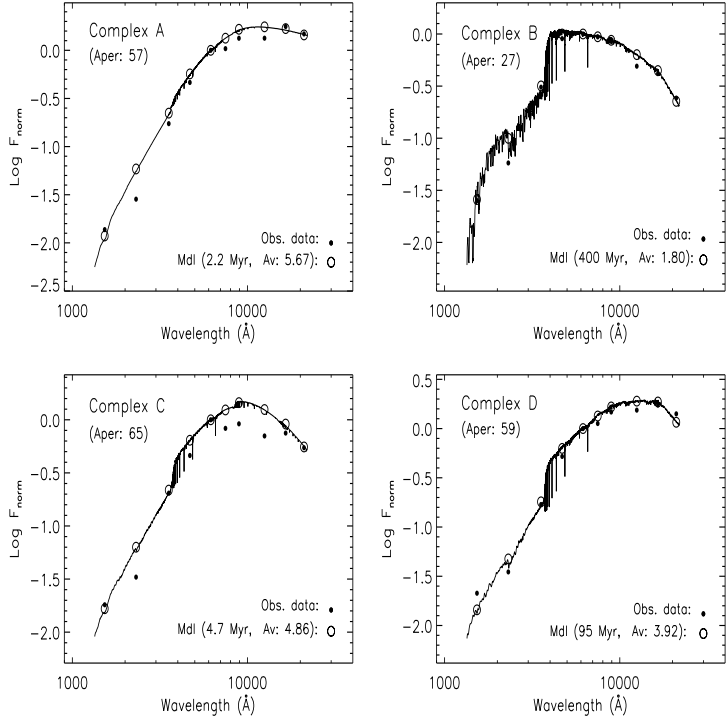}
\caption{The observed SEDs for complexes A, B, C and D (solid circles) are compared 
         with the 
         best-fit model SED (open circles) and the corresponding full resolution model 
         spectrum (solid line). The derived age and extinction are given in each plot.}
\label{fig:bestfit_seds_1}
\end{center}
\end{figure*}

\begin{figure*}[t]
\begin{center}
\includegraphics[width=0.6\textwidth]{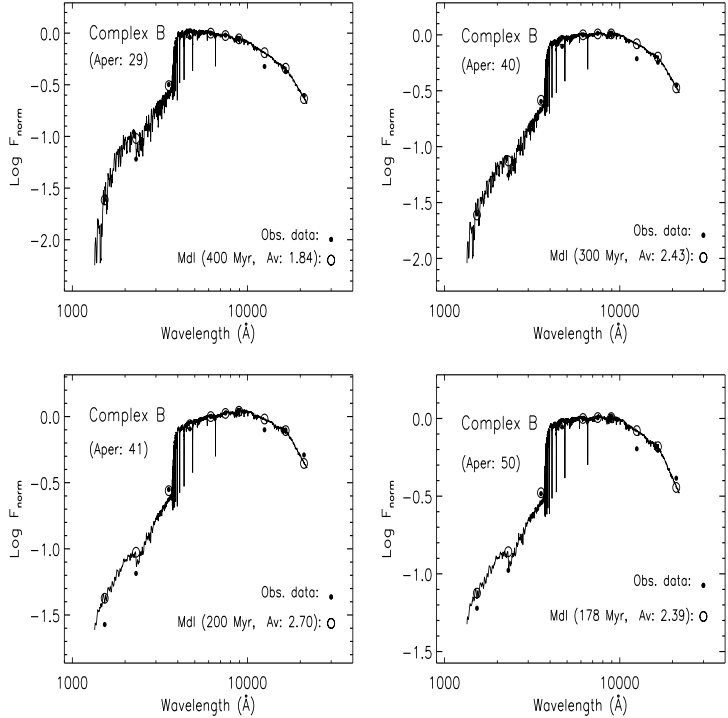}
\caption{Similar to Fig. \ref{fig:bestfit_seds_1}, but for some apertures located inside  
         complex B.} 
\label{fig:bestfit_seds_2}
\end{center}
\end{figure*}

Table \ref{tab:regs} lists the results obtained using the fitting method for apertures 
coincident with the high surface brightness complexes defined by \citet{OConnell78}. 
This table also shows  spectroscopic ages and extinctions of star clusters 
located inside the corresponding apertures.
Except for the complex B, the rest of the bright complexes have spatial
extensions similar to or smaller than the aperture size, and hence the derived ages
from aperture fluxes are the mean age of the complexes.
The complex B encloses 8 apertures, with the aperture 27 being the brightest.
Each of the complexes contain several compact star clusters. We used the high-resolution
HST images to estimate the contribution of the clusters to the aperture fluxes,
and found that the clusters contribute less than 1\% to the disk apertures,
whereas their contribution is less than 10\% for the nuclear apertures (see \S6). 
Hence, ages derived from the aperture fluxes do not represent those of the clusters,
instead they are representative of the field stellar populations of M82.
The SEDs of the best-fit models for complexes A, B (aperture 
27), C and D are shown in Figure \ref{fig:bestfit_seds_1}, for selected apertures of 
complex B in Figure \ref{fig:bestfit_seds_2}, and for complexes E, F, G 
and H in Figure \ref{fig:bestfit_seds_3}.

\subsection{Nuclear complexes: A, C, D and E}
Complexes A, C, D, and E are located in the nuclear star-forming region. 
This part of the galaxy shows strong $H\alpha$ emission \citep{Ohyama02}. The presence 
of ionized gas points to a population of stars of a few million years 
\citep{Rieke80,Satyapal97,Forster03}. More recently \citet{Konstan09} found  possible 
Wolf-Rayet features in the spectra of clusters located in these complexes, confirming 
that the clusters of the nuclear complexes are young.

Complex A is the core of the nuclear star-burst and is the brightest complex. 
This is the most 
studied complex of M82. Surrounding the complex A are complexes C, D and E. 
Each one of these 
complexes are resolved into several tens of CSCs in the {\it HST} images \citep{OConnell95}. 
Our aperture 57 encloses the bright core of the complex A, for which we obtained an age of 
$\sim2$~Myr ($log\;age=\;6.31\pm0.05$) and $A_v=5.7$~mag.

For complexes C and E, we derived ages of around 5 and 7~Myr ($log\;age=\;6.66\pm0.03$ and 
$log\;age=\;6.85\pm0.02$, respectively) with $A_v$ values of $\sim$5 mag. These derived 
parameters are consistent with the previous results \citep{Forster03}. For complex D, which is 
directly to the north of complex A, we derived an age of $\sim$95~Myr ($log\;age=\;7.97\pm0.07$). 
This age is clearly larger than the age of the nuclear starburst and probably this belongs 
to the disk and it is seen projected on top of the nucleus of the galaxy.

\subsection{Complex B}
Complex B is the most studied complex outside the nuclear zone. It lies around a kiloparsec 
to the north-east of the nucleus. In the blue band images this complex is even brighter than 
the nucleus. Complex B spans over a large area of the disk of M82, and is the largest of 
the named complexes. Sometimes, it is subdivided into two parts, the B1 containing the bright 
zones to the far east and B2 containing relatively fainter zones to the west \citep[][]
{DeGrijs01}. Using long-slit spectroscopy, \citet{Mayya06} obtained an age of 500~Myr for 
this complex. There are $\sim40$ CSCs  in this complex \citep{Mayya08}. Recent spectroscopic 
studies have obtained ages for some of these clusters. \citet{Smith07} found that stellar 
clusters located in this area cover an age interval of 80--270~Myr, peaking at 150~Myr. 
\citet{Konstan09} determined that star clusters located in this complex are 80--200~Myr old. 

In our analysis, complex B is covered by 8 apertures. We have listed these apertures in 
Table \ref{tab:regs} as function of the distance to the center of the galaxy. Figure 
\ref{fig:bestfit_seds_2} displays the best-fit SEDs for selected apertures located inside this 
complex. Apertures 27 and 29,  which lie to the extreme east contain the brightest part of this 
complex, for which we derived ages of $\sim$400~Myr ($log\;age=\;8.60\pm0.06$, and 
$log\;age=\;8.59\pm0.06$, respectively). The age derived for the aperture  50 -- which lies 
on the extreme west of this complex --  is $\sim$180~Myr ($log\;age=\;8.25\pm0.07$). We note 
that the ages of the western apertures are systematically younger reaching values of around 
200~Myr at the extreme west (see Table \ref{tab:regs}).

\subsection{Complexes F, G and H}

\begin{figure*}[t]
\begin{center}
\includegraphics[width=0.6\textwidth]{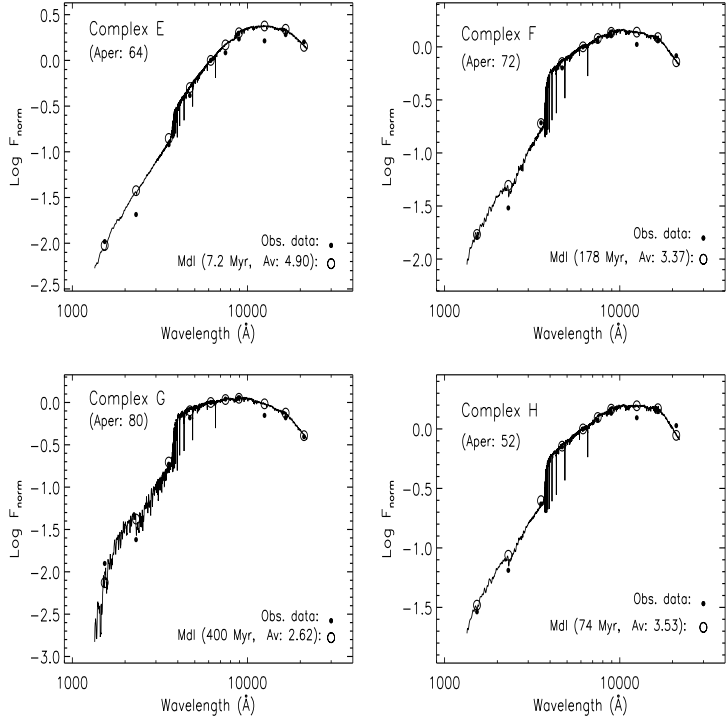}
\caption{Similar to Fig. \ref{fig:bestfit_seds_1}, but for complexes E, F, G and H.} 
\label{fig:bestfit_seds_3}
\end{center}
\end{figure*}

Complexes F and H lie in the transition zone between the nucleus and the disk. Each of these 
complexes is associated with a very luminous and massive stellar cluster. Complex F 
is located 
on the south-west of the nucleus (aperture 72), whereas complex H is at the north-east of 
the nucleus (aperture 52). Complex G (aperture 80) is located in the disk on the west side. 
This latter complex was not included in previous spectroscopic studies of M82 probably 
because it is very diffuse. 

\citet{Smith01} performed a spectroscopic study of the stellar cluster of complex F. They found 
that this cluster is around 50~Myr old. \citet{Bastian07} found substructures 
within the cluster, and in the surrounding region, which was interpreted as due to 
differential extinction across the face of this complex, with some regions having $A_v>2.5$ mag. 
They estimated 
an age of 60~Myr. \citet{Konstan09} analyzed two clusters in this area (slits 43.2 and 51(F)). 
They concluded that the age of cluster F is 40--80~Myr (in accordance with result obtained 
by \citet{Smith01} and \citet{Bastian07}), whereas cluster 43.2 is 100--320~Myr old. 

Complex H is dominated by one of 
the most massive star cluster of M82 ($\sim2.6\times10^5\;M_\odot$, \citet{Mayya08}) which is 
also known as cluster B2-1. 
Using {\it HST} images, \citet{Konstan08} found a peculiar extinction pattern associated 
to this cluster,  
complicating its 
age determination. In a more recent analysis \citet{Konstan09} determined that the cluster of 
complex H is 180--200~Myr old. 

Figure \ref{fig:bestfit_seds_3} displays the best-fit models for these complexes. For 
complex F we found a stellar population of around 180~Myr ($log\;age=\;8.25\pm0.07$). 
For the population of complex G we determined an age of $\sim350$~Myr. For the 
complex H, we obtained an age of $\sim75$~Myr ($log\;age=\;7.87\pm0.05$), which is 
younger than the star cluster B2-1.

\begin{deluxetable}{cccccccccl}
\rotate
\tablecolumns{10}
\tabletypesize{\scriptsize}
\tablecaption{Age of the Stellar population and extinction found in apertures related 
              with complexes A, B, C, D, E, F, G and H \label{tab:regs}}
\tablewidth{0pt}
\tablehead{
\colhead{Complex} & \multicolumn{3}{c}{Our Results (Field Stars \& Clusters)} & \colhead{} & \multicolumn{3}{c}{Konstantopoulos et al. 2009 (Clusters)} & \colhead{} & \colhead{Other authors} (Clusters)\\
\cline{2-4} \cline{6-8} \\
\colhead{ } & \colhead{Apertures} & \colhead{Age (Myr)} & \colhead{A$_v$}  & \colhead{} &  \colhead{Slit} & \colhead{Age (Myr)} & \colhead{A$_v$}  & \colhead{} & \colhead{}} 
\startdata
A & 57 &  2  & 5.67 & &  78.1,78.2  &   30,20    &   1.20,3.45    & &       6.4 Myr and E(B-V)=1.35 $^{1}$ \\
\\
B & 27 & 403 & 1.80 & &    131      &     80     &      1.08      & & 80--270 Myr (peak at 150 Myr) $^{3}$ \\
  & 29 & 396 & 1.84 & & 125,126,131 & 140,210,80 & 1.17,0.85,1.08 & &                                       \\
  & 32 & 366 & 1.82 & &     ...     &     ...    &       ...      & &                                       \\
  & 35 & 357 & 2.21 & &    113      &      110   &      1.67      & &                                       \\
  & 40 & 316 & 2.43 & &     ...     &      ...   &       ...      & &                                       \\
  & 41 & 197 & 2.70 & &     ...     &      ...   &       ...      & &                                       \\
  & 45 & 263 & 2.16 & &     ...     &       ...  &       ...      & &                                       \\
  & 50 & 180 & 2.39 & &    97,98    &  160,160   &    0.90,0.50   & &                                       \\
\\
C & 65 &  5  & 4.86 & &     58      &      30    &      1.12      & &                                       \\
\\
D & 59 & 95  & 3.92 & &    67.4     &       7    &      2.12      & &                                        \\
\\
E & 64 &  7  & 4.90 & &  67.1,67.2  &      7,7   &   2.47,2.17    & &                                        \\
\\
F & 72 & 180 & 3.37 & &  43.2,51(F) &    130,70  &   1.02,2.42    & &   60 -- 80 Myr and Av$>$2.5 mag $^{2}$ \\
  &    &     &      & &             &            &                & &                          50 Myr $^{1}$ \\
\\
G & 80 & 356 & 2.62 & &             &            &                & &                                         \\
\\
H & 52 & 75  & 3.53 & &    91(H)    &     200    &       2.40     & &         346 Myr and E(B-V)= 0.97 $^{1}$ \\
\enddata
\tablerefs{
(1) Smith et al. 2006;
(2) Bastian et al. 2007;
(3) Smith et al 2007.
}
\end{deluxetable}

\section{Spatially Resolved Star Formation History and Extinction}

In this section, we use the age and extinction of all the apertures to infer 
the spatially resolved star formation history in the disk of M82 in scales of 
$\sim200$ pc.

\begin{figure}[h]
\epsscale{1.05}
\plotone{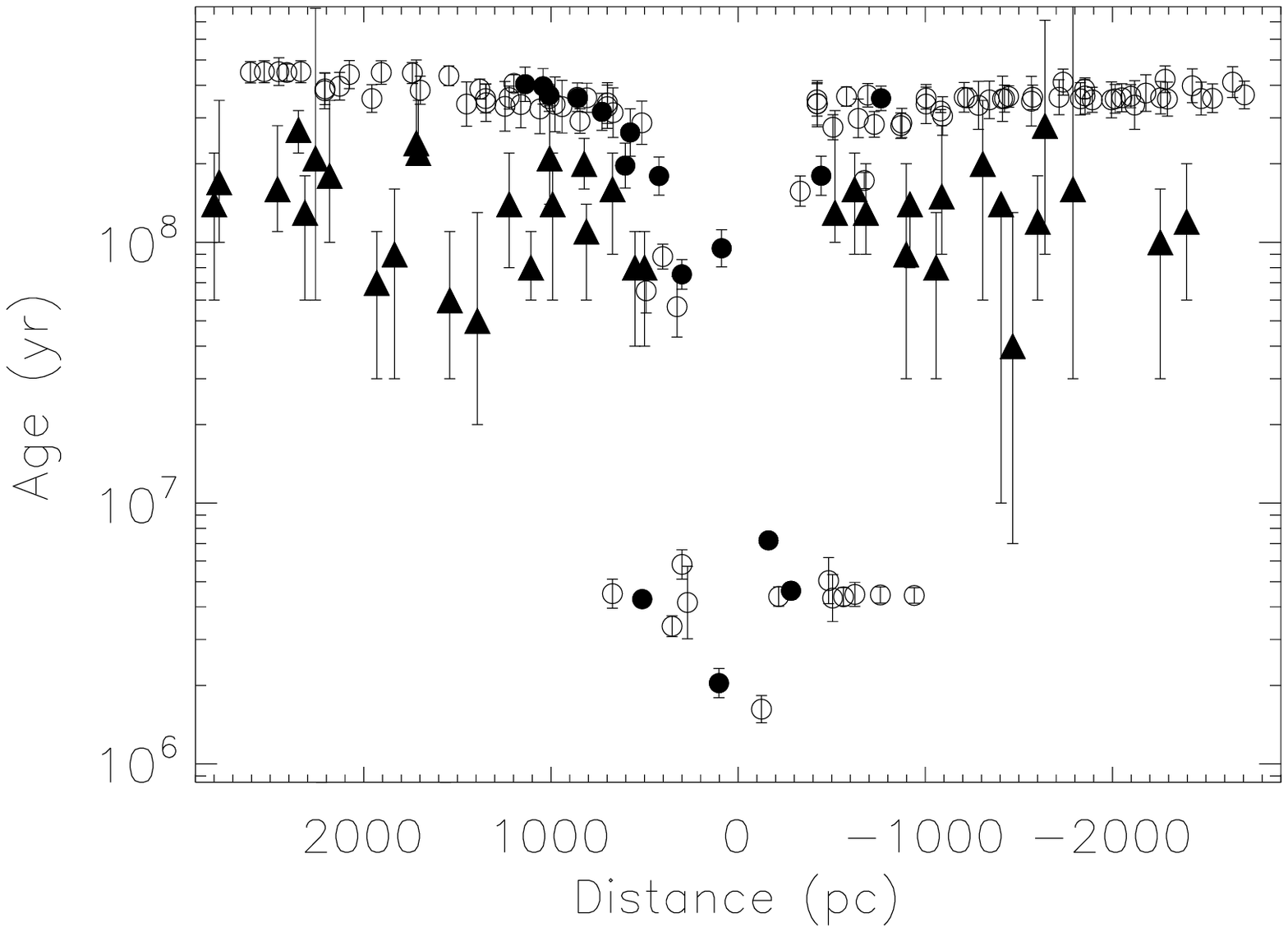}
\caption{Derived ages of the dominant stellar population within apertures of 
         $5\arcsec$ radius as a function of the distance 
         to the galactic center (positive distances correspond to the north-east 
         direction). Circles mark the results for the selected apertures, with the 
         filled circles marking the complexes defined by \citet{OConnell78}. 
         Ages of the clusters from \citet{Konstan09} are denoted by triangles.}
\label{fig:age_radio}
\end{figure}

\begin{figure}[h]
\epsscale{1.05}
\plotone{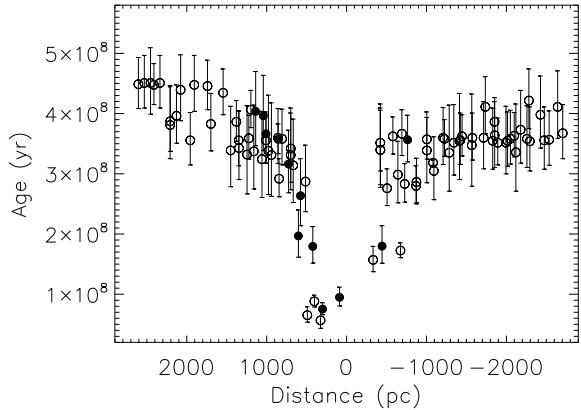}
\caption{As in previous Figure, but with the vertical axis on a linear scale 
         to enable better visualization of a slightly outside-in age gradient 
         in the disk of M82.}
\label{fig:age_radio_b}
\end{figure}

Figure~\ref{fig:age_radio} shows the distribution of age as a function of the distance 
to the center of the galaxy. We distinguish those apertures belonging to complexes 
A to H by denoting them with filled circles. Two distinct epochs of star formation can 
be easily inferred from this plot: one younger than 10~Myr, exclusively located in the 
central 0.5~kpc radius, and the other older than 100~Myr, distributed along the whole disk 
of the galaxy. 
In Figure~\ref{fig:age_radio_b}, we display
the age of the apertures as a function of the distance using linear scale.
A trend can be seen in this plot in such a way that the ages decrease
from around 450~Myr at $\sim$3 kpc to 300~Myr at 0.5~kpc on the eastern side
(positive distances), and from 400~Myr at 3~kpc to 300~Myr at 0.5~kpc on the western side. 
In the transition zones between the disk and the nucleus, 
the ages decrease rapidly from $\sim300$~Myr to $\le100$~Myr.
The observed trend may correspond to a smooth age gradient in the disk. However,
the uncertainties in the derived ages in this work, do not permit a clear
answer on this.

Another important thing to note in Figure~\ref{fig:age_radio_b} is the similarity 
in ages of the high surface brightness named complexes and the surrounding disk. 
This indicates that the stellar populations in the field, as well as in the 
bright complexes, were formed in the same event. Implications of these results will 
be discussed in \S7.

\citet{Konstan09} used multi-object spectroscopy to obtain ages and extinctions for 49 
relatively bright clusters, covering up to a radial distance of 3~kpc. Their ages are 
displayed in Figure~\ref{fig:age_radio} using filled triangles. They found a mean age 
of 150~Myr for disk clusters, with no 
noticeable age gradient. Hence, these clusters seem to be systematically younger than 
the field populations as traced by our apertures. This difference makes the clusters 
slightly bluer than the surrounding disk as was noted by \citet{Mayya08}. The
implications of these differences are discussed in \S7.

\begin{figure}[h]
\epsscale{1.05}
\plotone{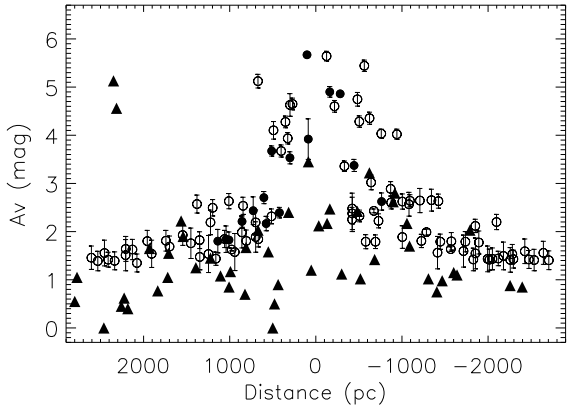}
\caption{The visual extinction within apertures of $5\arcsec$ radius as a 
         function of the distance to the galactic center (positive distances 
         correspond to the north-east direction). The central part of the galaxy 
         is more extincted, $A_V>3.5$ mag. Circles describe our results, filled 
         circles are related to named complexes, and triangles represent $A_V$ in
         clusters obtained by \citet{Konstan09}.}
\label{fig:extin_radio}
\end{figure}

Several studies have found an extinction gradient in the central kiloparsec of M82 with 
the extinction increasing towards the nucleus \citep{Waller92}. Most of these studies 
are based on nebular line ratios and hence sample only the central part of M82. Our 
method allows mapping of the extinction in the entire disk. In Figure~\ref{fig:extin_radio}, 
we show the radial distribution of extinction. A gradient of extinction can be seen in 
the disk on both sides of the nucleus, with the $A_v$ increasing smoothly from 1.5~mag 
at 3~kpc to 2.8~mag at 0.5~kpc. For the nuclear apertures, we found extinction ranging 
between 3--6~mag, which agrees well with the values obtained by \citet{Waller92} using 
nebular lines. On the other hand, the star clusters studied by \citet{Konstan09} present 
extinction values that are systematically $\sim1$ mag lower than our values for selected 
apertures (see Table \ref{tab:regs}).

\subsection{The underlying old disk}

It is important to note that we could reproduce the SEDs of all the disk apertures
using a single stellar population of age $\sim400$~Myr, without the need to add an
SSP of several gigayears of an underlying old disk.
These results are consistent with the disk-wide star-formation episode proposed
by \citet{Mayya06}, in which more than 90\% of stellar mass in the disk was
formed.
The derived aperture ages are consistent also with that inferred by \citet{Davidge08}, 
and are marginally younger than that proposed by \citet{Mayya06}.

\begin{figure}[ht]
\epsscale{1.05}
\plotone{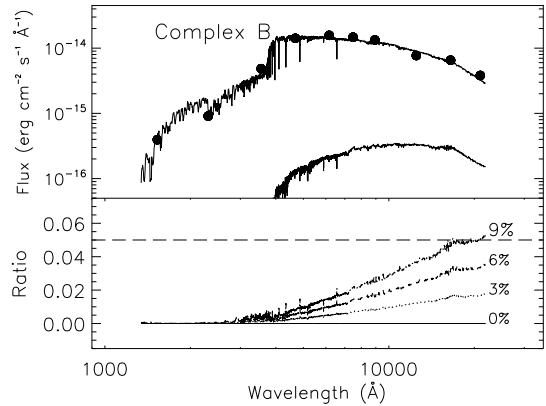}
\caption{The observed SED (circles) of an area of complex B (aperture 27) is plotted in the
         upper panel, as well as the SED of the best-fit SSP (solid line fitting the dots) 
         and the SED of the old population model (solid line) with 9\% in mass of the best-fit 
         model. The lower panel shows the ratio ($F_{Old-Pop}/F_{Best-Fit}$) for different 
         fractional masses (indicated as percentage for each curve) of the old population. 
         For the {aperture} plotted, a mass$>$9\% would have been noticed in the K-band photometry.}
\label{fig:add_mass}
\end{figure}

\begin{figure}[ht]
\epsscale{1.05}
\plotone{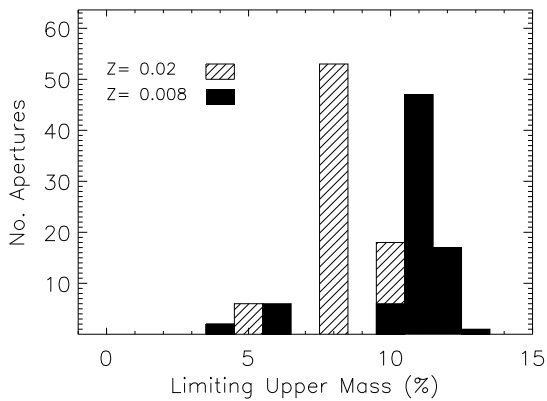}
\caption{The histogram shows the limiting upper masses of old population that can be 
         added to the best-fit SSP of apertures located in the disk of M82. The 
         mean value is 9\% for solar metallicity models and 11\% for metal-poor models.}
\label{fig:hist_old_mass}
\end{figure}

In this section, we check whether our multi-band SEDs are consistent with less than
10\% of mass in old underlying disk stars.
In order to do this, we performed new fits between the observed SEDs and new models. 
The new models were calculated by adding an SSP of 4 Gyr old to the best-fit SSP discussed 
in previous sections. We assumed that the old component shares the same reddening as the 
best-fit SSP. We carried out two tests: one using solar metallicity for the old 
component and a second test using a metal poor model (Z= 0.008) for the old component. 
The mass of the old population (old-pop) model is varied from a minimum of 0\% in mass 
of the best-fit SSP in steps of 1\% until the flux of the new model exceeds the observed 
flux plus the estimated error in the K band. We have chosen the K-band, because the old 
population is expected to have a maximum contribution in this band. The analysis performed 
for an area of complex B (aperture 27) using solar metallicity is illustrated in 
Figure \ref{fig:add_mass}. The upper plot displays the observed SED (dots), the SED of 
the best-fit model (solid line fitting the dots) and the SED of the old population (solid 
line) when it has a 9\% in mass of the best-fit model. The lower plot shows the comparison 
between the flux of the old-pop model (with different percentage of mass) and the flux of 
the best-fit model ($Ratio= F_{Old-Pop}/F_{Best-Fit}$). It can be inferred that the old 
population {with solar metallicity} at the most contains 9\% of the stellar mass of the 
best-fit population inside aperture 27. In the case of a metal-poor model, the upper 
limit in mass is 11\%. More massive old stellar population would have increased the K-band 
flux to values above that observed. The upper mass limit of the old population obtained 
for aperture 27 is typical for the rest of the disk apertures, as is illustrated in the 
histogram of Figure~\ref{fig:hist_old_mass}. The mean upper mass limit is 9\% and 
11\% depending on whether the old population is assumed to have solar or sub-solar metallicity,  respectively. Thus, the observed multi-band SEDs of individual apertures are consistent 
with the result obtained by \citet{Mayya06} where the underlying old disk, if present, 
is very insignificant in M82.

\section{Cluster Formation Efficiency}

From the analysis of the previous sections, it is clear that the observed aperture fluxes 
of both the disk and nuclear regions of M82 are dominated by a single generation of stars 
modeled as a simple stellar population. M82 contains a rich population of CSCs, with 
each CSC containing a single generation of stars \citep[][]{Mayya08}. So, it is interesting 
to investigate whether the CSCs and the field stellar populations formed from the same 
episode. 

In Figure~\ref{fig:age_radio}, we have compared the ages derived using our aperture fluxes 
to the ages of stellar clusters located within the apertures. The disk clusters have a 
mean age of 150~Myr, whereas most of the field populations surrounding these clusters have 
ages between 300--400~Myr. Thus, clusters seem to be systematically younger than the 
surrounding field populations by around 200~Myr. In the disk-wide starburst model proposed 
by \citet[][]{Mayya06}, the star formation episode in the disk lasted for around 300~Myr. 
Thus, in spite of the systematic differences in mean ages, it is still possible that both 
the clusters and the field stars formed from the same episode. On the other hand, in the 
nuclear region, both clusters and stars in their vicinity have similar ages, both populations 
being part of the on-going nuclear starburst.

Another question of interest is to find out whether clusters dominate the observed fluxes 
at spatial scales of 180~pc. In order to answer this question, we compared the total fluxes 
of all clusters located physically within an aperture of $5\arcsec$ radius with the integrated 
fluxes within that aperture.
\begin{figure}[t]
\epsscale{1.05}
\plotone{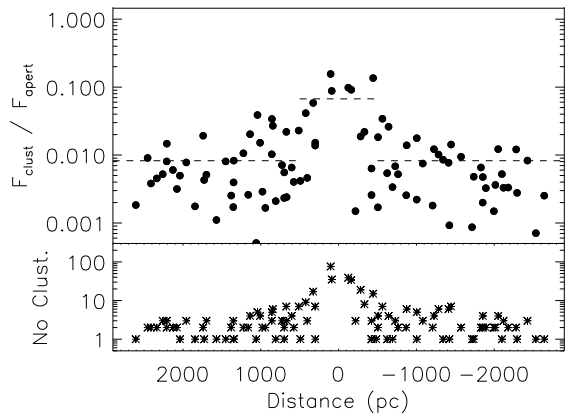}
\caption{The contribution to the total flux by the stellar clusters as a function of the 
         distance to the galactic center (each point corresponds to an aperture ). The upper 
         plot shows that the mean contribution in the nucleus is $\sim7\%$, whereas in the 
         disk it is around 0.7\% (dashed lines). The lower plot shows the number of star 
         clusters inside each aperture (as in previous figures, positive distances 
         correspond with the NE).}
\label{fig:ncluster_v}
\end{figure}
Typically, nuclear apertures contain more than 10 compact clusters with the aperture 57 
(complex A) containing as much as 76, whereas disk apertures contain less than 10 clusters 
and  several apertures do not have any catalogued cluster inside. 
We calculated the 
fluxes of the field population using the apertures defined in Figure~1 on the
F555W image obtained from  {\it HST/ACS} observations \citep{Mutchler07}.
The F555W band cluster fluxes were taken from \citet{Mayya08}.

Figure \ref{fig:ncluster_v} displays in the upper plot the comparison between the observed 
flux and the fluxes emitted by stellar clusters in the selected apertures. The lower plot 
shows the number of star clusters inside each aperture. Clusters contribute clearly more 
to the aperture fluxes in the nucleus as compared to the disk, 7\% vs. 0.7\%. These results 
show that field stars, not the clusters, are emitting most of the observed flux at scales 
of around 180 pc, even in 
areas where there is strong star formation. Similar results are obtained with the F435W and 
F814W images from {\it HST/ACS}. There are apertures located inside the 500 pc radius with ratios 
below $\sim$2\%. These apertures seem to belong to the disk far away from the center of 
the galaxy but due to the high inclination of the galaxy are seen projected on to the nuclear 
region. The derived ages and extinctions of these regions confirm that interpretation.

The flux fraction obtained above is directly related to the fraction of stellar mass 
in compact clusters as compared to the mass in all field stars, provided the
clusters and field stars are of the same age. For the nuclear region this assumption
holds, and hence on an average, 7\% of the stars in the nuclear region of M82 form 
in clusters. As these nuclear populations are younger than 10~Myr, the observed
fraction corresponds to the cluster formation efficiency. Curiously, 
this estimate compares very well with the cluster formation efficiency 
of $8\pm3\%$ determined by \citet{Bastian08} in star-forming galaxies.

On the other hand, the disk clusters are younger than the surrounding
field stars. Given that the fluxes of stellar population decrease with time
due to passive evolution, the observed flux fraction of 0.7\% is an upper limit to the
fraction of mass in cluster stars.

\citet{Meurer95}, working with the {\it HST} UV images of nine starburst galaxies with 
several bright star clusters, found that 20\% of the total UV luminosity is emitted by 
the clusters. It is most likely that the typical age of star formation in his sample 
galaxies is only around 10~Myr. Our values for the nucleus of M82 reaches 20\% in one 
case with the rest having values below those obtained by \citet{Meurer95}. So it seems 
that CSCs contribute $<20\%$ to the flux in young star forming regions decreasing to 
less than one percent when they are several hundreds of million years old.

\section{Discussion and Implications}

We start our discussion addressing the nature of complex B, which has drawn a lot of 
attention in recent years \citep{DeGrijs01,Smith07,Konstan09b}. Often this complex 
is referred to as fossil starburst 
\citep{DeGrijs01}. There was also the suggestion that the brightness of this complex 
is due to windows in the dust distribution \citep{Konstan08}. Results obtained with 
our fitting method show that ages and extinctions of the bright named complexes are 
similar to the surrounding relatively fainter areas (see Figures~\ref{fig:age_radio} 
and \ref{fig:extin_radio}), suggesting that the high brightness of the named complexes 
is due to their higher star formation rate with respect to the surrounding disk, 
and not because of lower extinction or younger age. Particularly, apertures belonging 
to complex B do not show significant lower extinction as often claimed \citep{Konstan08}. 

Ages of CSCs obtained using spectroscopy by \citet{Konstan09} 
are in the range of 100-300~Myr with the mean age of 150~Myr. Thus clusters seem to 
be systematically younger than the surrounding disk stars. However, it is important 
to point out that only around 50 of the 653 known clusters \citep{Mayya08} 
were selected for spectroscopic 
observations and hence the derived ages may not be representative of the entire cluster 
population. Given that the cluster luminosity decreases with age, the brightest clusters 
are expected to be the youngest. Hence it is not strange that the derived ages of the clusters 
are systematically younger. 
Complex H, which lies in the transition zone between the nucleus and disk,
offers an interesting case, where the compact cluster is older than the
field population surrounding it.

\citet{Mayya06} found that the entire disk participated in a star-forming episode 
that started following the interaction with M81 around 500 Myr ago. 
Ages of the dominant population 
found in our study are consistent with that idea. However, we find in the disk of 
M82 a systematic trend for the external zones to be marginally older as compared 
to the inner zones. Considering that our ages are luminosity weighted, this trend 
suggests either that the star-forming episode was not simultaneous in the entire 
disk but it is systematically younger 
in the inner regions, or that the entire disk started forming stars at the 
same time, but the star-forming episode stopped systematically at earlier epochs 
in the outer regions. Study of resolved stellar populations by \citet{Davidge08} 
supports the latter idea over a much extended disk. 

We commented that the mean SEDs of central, north-east and south-west parts of the 
disk display differences (see Fig. \ref{fig:seds}). 
The age and extinction distributions along the disk of the galaxy can explain these 
differences. The visual absorption found on each side of the galaxy is similar, the mean 
$A_{V}$ of the NE is 1.6 mag and the mean $A_{V}$ of the SW is 1.8 mag. However, 
the population 
in the south-west is slightly younger: the oldest population of the SW is around 400 Myr old 
whereas in the NE side the oldest population is $\sim$450 Myr old (see Figs. \ref{fig:age_radio_b} 
and \ref{fig:extin_radio}). This behavior explains the bluer appearance of mean SED of the 
south-west part. On the other hand, we have pointed out that the mean SED of the central part 
is redder than the mean SEDs of the NE and SW regions. In spite of the very young ages found 
in the central parts of the galaxy, the high optical extinction ($A_{V}$ around 5 mag, see 
Fig. \ref{fig:extin_radio}) found in this part produces a mean SED which looks very 
reddened.

There is very little evidence for the existence of stars formed previous to the known 
star-forming episode. In the star-formation model proposed to explain the properties 
of the disk, \citet{Mayya06} found a maximum contribution of around 10\% of the total 
mass in stars older than 1 Gyr. In our study, we have analyzed this fraction in every 
aperture. We found that old stars do not contribute more than 11\% in majority 
of the disk apertures. Thus, we reiterate that the disk of M82 formed almost all of 
its stars after the interaction with M81.

Young compact clusters contribute $\sim$7\% to the nuclear aperture fluxes, with the 
corresponding contribution being $\sim$0.7\% for the relatively older disk apertures. 
Given that the clusters and the surrounding regions  share similar age and extinction, the 
above fraction is an indication of the efficiency of cluster formation and its dissolution. 
Hence for young nuclear clusters, the efficiency on an average is around 7\%. The observed 
relatively lower fractions for the disk regions may indicate that on the time scale of a few 
hundred million years many clusters are already dissolved.

\section{Summary}

In this work,
we analyzed the star formation history of M82 over spatial scales of $\sim180$~pc using 
the broad-band SEDs covering the FUV to NIR bands. We constructed 117 SEDs, where each 
point of the SED is obtained by carrying out photometry in apertures of 5$^{\prime\prime}$ 
radius over the entire face of M82 within 3~kpc radius. These SEDs were fitted with 
synthetic SSPs in wide range of reddening and ages. With this technique, we confirmed 
the following two facts that were already known: 
(1) the stellar populations in the nuclear zone are less than 10~Myr old and are
heavily reddened ($A_v=$2--6~mag),
(2) the stellar populations in the disk have ages between 100--450~Myr.

The use of spatially resolved SEDs has allowed us to establish, 
for the first time, that the interaction-driven disk star formation 
was not restricted to a few bright zones, but occurred everywhere in the
disk over spatial scales as small as 180~pc.
Each of these SEDs is consistent with the idea that more than 90\% of the stellar
mass of the disk is formed as part of the disk-wide star formation episode, 
and containing only a small fraction of stars older than 1 Gyr.

We found that the bright named complexes have similar ages and reddening as 
that of the surrounding low-surface brightness parts of the galaxy: complexes 
within the starburst region (complexes A, C, E) have ages less than 10~Myr and 
the complex B in the disk has age 200--400~Myr, and the complexes in the 
intermediate zone (F and H) having ages between 50--200~Myr. This implies that 
the named complexes are brighter just because of higher amount of star formation, 
not because they are younger or have lesser reddening. Ages of compact stellar 
clusters in the disk derived using spectroscopy ($\sim150$~Myr) are systematically 
smaller than the mean age of the field stellar populations ($\sim350$~Myr). However, 
the difference in the age is of the same order as the duration of the disk-wide 
star formation episode, and hence both populations could belong to the same episode.

In the nuclear regions, around 7\% of all the stellar mass resides in the cluster
stars, whereas that fraction is less than 1\% in the disk. Given that the
disk populations are systematically older than the nuclear populations,
the difference in the fractional mass in cluster stars may be due to disruption
of clusters as they evolve.

The results obtained in this study are encouraging from the point of view
of extending the technique adopted in this work to other nearby galaxies.
Thus star formation history over a few hundreds of parsec scale can be 
derived by making use of the already available multi-band archival images
such as {\it GALEX}, {\it SDSS}, and {\it 2MASS}.

\vspace{1cm}
The authors are pleased to thank R. Terlevich, E. Terlevich, A. Bressan, O. Vega, and 
A. Luna for fruitful discussions and helpful suggestions. We also thank the anonymous 
referee for many useful comments that have lead to an improvement of the original
manuscript. This work is partly supported by CONACyT (Mexico) research grants 58956-F 
and 49942-F.

Some of the data presented in this paper were obtained from the Multimission Archive at 
the Space Telescope Science Institute (MAST). STScI is operated by the Association of 
Universities for Research in Astronomy, Inc., under NASA contract NAS5-26555. Support 
for MAST for non-HST data is provided by the NASA Office of Space Science via grant 
NNX09AF08G and by other grants and contracts.


\end{document}